# Machine learning potential-driven prediction of high-entropy ceramics with ultra-high melting points

Hong Meng, Yiwen Liu, Hulei Yu*, Lei Zhuang, Yanhui Chu*

School of Materials Science and Engineering, South China University of Technology, Guangzhou, 510641, China

*Corresponding author.
*E-mail address:* huleiyu@scut.edu.cn (H. Yu); chuyh@scut.edu.cn (Y. Chu)

## Abstract

Developing high-entropy ceramics (HECs) with ultra-high melting points ($T_m$) is crucial for their applications in ultra-high-temperature environments. However, related research has seldom been reported. Here, taking high-entropy diborides (HEBs) as an example, we develop a data-driven method to efficiently explore HEBs with ultra-high $T_m$ via transferable machine-learning-potential-based molecular dynamics (MD). Specifically, a moment tensor potential (MTP) for HEBs with nine transition metal elements of group IVB, VB, and VIB is first constructed based on unary and binary diborides. Further studies on the performance of our constructed MTP have confirmed its remarkable accuracy, transferability, and reliability across both equimolar and non-equimolar HEB systems. $T_m$ of HEBs are then accurately simulated through MD simulations based on the constructed MTP, and 24 features are simultaneously collected to enable reliable machine learning training. Five descriptors with the gradient boosting regression model are derived as the optimal combination for accurate $T_m$ predictions in HEBs with genetic algorithms. Based on our established model, $T_m$ of 32563 HEBs are eventually determined, achieving the maximum $T_m$ of 3688 K in $(Ti_{0.1}Zr_{0.1}Hf_{0.6}Ta_{0.2})B_2$. The work presents a feasible approach to develop HECs with ultra-high $T_m$.

## 1. Introduction

High-entropy ceramics (HECs), a class of inorganic compound solid solutions with one or more Wyckoff sites shared by no less than four principal elements, have received growing interest from the materials community in extreme environment research [1-5]. Since the first report of HECs in 2015 [1], a diverse range of HECs, like high-entropy oxides [1,6-8], high-entropy diborides (HEBs) [2,9-11], and high-entropy carbides [3,4,12,13], have been developed. Beneficial from the “cocktail effect”, these materials have been found to exhibit exceptional properties suppressing their parent ceramics, such as better thermal stability [1,7,8,11], enhanced oxidation and ablation resistance [10,12,15,16], superior thermo-mechanical properties [2,11,17], higher thermal insulation [8,14,18], and improved radiation resistance [19,20], making them highly anticipated candidates in the extreme environment application, such as high temperature, high pressure, and strong erosion and radiation. Among these distinct behaviors, melting points ($T_m$) are of great importance to the high-temperature applications of HECs. However, the enormous composition space has posed a great challenge in developing HECs with ultra-high $T_m$. Therefore, it is urgent to develop an efficient and reliable approach to search for HECs with ultra-high $T_m$.

Stemming from the recent advancement of machine learning (ML) algorithms and data science, high-throughput predictions based on ML methods have become the most effective solution to material discovery in HECs [21-23]. Successful applications of ML predictions have been reported on the phase formation ability of various HECs in recent studies [6, 24-26]. Nevertheless, the data-driven predictions of $T_m$ are severely limited due to insufficient $T_m$ data of HECs collected from laborious and costly experiments. To this end, molecular dynamics (MD) is deemed the ideal method to efficiently simulate $T_m$ of HECs [27,28]. However, the absence of reasonable interatomic potentials for HEC systems presents a major obstacle to their wider implementation. Recent progress has identified machine learning potentials (MLPs) as a powerful tool to achieve MD simulations with the accuracy of *ab initio* methods while beyond their efficiency, length, and timescale limitations. Up to now, efforts have been successfully made to investigate various properties of different ceramics with MLPs, such as $ZrB_2$ [29] and $TiO_2$ [30]. However, the low transferability of MLP constructed by traditional strategy [31,32] significantly limited their wide applications in HECs. Therefore, attempts to utilize MD simulations with MLPs to facilitate data-driven discovery of HECs with ultra-high $T_m$ have been seldom reported.

In this study, taking HEBs as an example, we advance a data-driven prediction on searching potential HEBs with ultra-high $T_m$ via transferable MLP-based MD simulations. Specifically, a transferable moment tensor potential (MTP) for HEBs with broad compositional applicability has been well-constructed according to our recently developed strategy [33]. As shown in **Fig.1**, the training dataset is effectively generated from unary and binary diborides with nine transition metal (TM) elements of group IVB, VB, and VIB through density functional theory (DFT) calculations. The accuracy and transferability of the trained MTP are then examined by both equimolar and non-equimolar HEB testing datasets with low root mean square errors (RMSEs) in both energies and forces. The reliability of our established MTP is further confirmed by a comparative analysis of the structural and mechanical properties between MD simulations with our established MTP and DFT calculations. To establish the ML model, the accurately calculated $T_m$ of 120 HEBs and nine diborides are collected from MLP-based MD simulations, and 24 descriptors are simultaneously computed to achieve reliable ML models for the $T_m$ prediction in HEBs. Optimal combinations of five descriptors are determined by ML approaches combined with genetic algorithms (GAs). Eventually, high-throughput screening on $T_m$ of equimolar and non-equimolar HEBs is carried out to search for potential HEBs with ultra-high $T_m$. The work offers a new way to accelerate the discovery of HECs with ultra-high $T_m$.

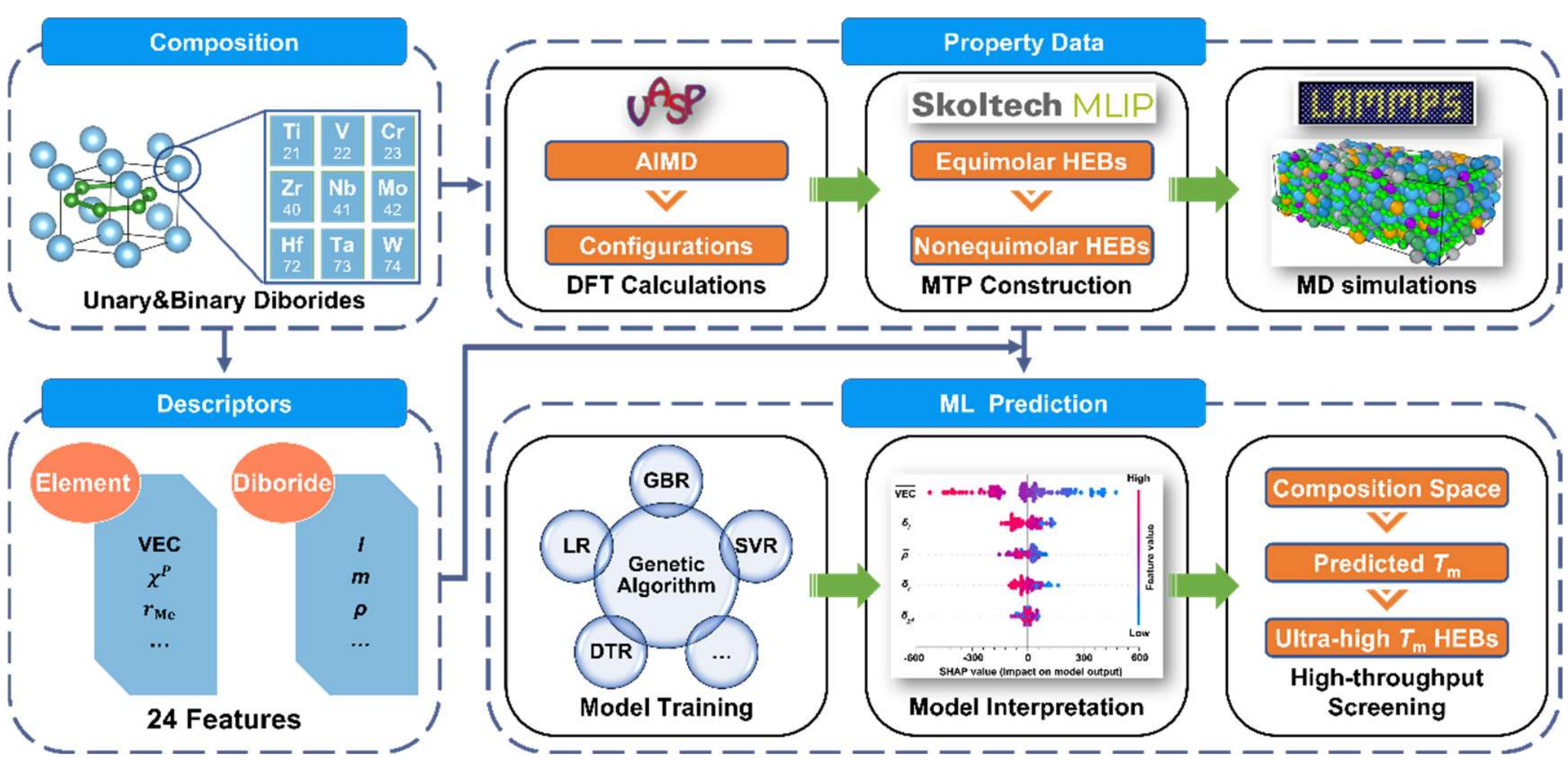

**Fig. 1** Schematic of ML predictions for HEBs with ultra-high $T_m$.

## 2. Computational details

DFT calculations were implemented in the Vienna *Ab Initio* Simulation Package (VASP) with the projector augmented wave method (PAW) [34] and the Pardew-Burke-Ernzerhof (PBE) [35] version of the generalized gradient approximation (GGA) [36].

The cutoff energy for the plane-wave basis of all calculations was set to 450 eV, and the convergence criteria for ionic and electronic steps were set to $10^{-3}$ and $10^{-4}$ eV, respectively. In addition, a grid spacing of 0.3 $Å^{-1}$ with the Γ-centered mesh was used in the Brillouin zone [37], and the spin polarization effects were taken into account. Based on the $TiB_2$ conventional cell with a space group of *P*6/*mmm*, 3 × 3 × 4, 2 × 2 × 4, 2 × 2 × 5, 2 × 2 × 6, 2 × 2 × 7, and 4 × 4 × 2 supercells were built to simulate unary (as well as binary and nonary), quaternary, quinary, senary, septenary, and octonary TM diborides, respectively. The special quasi-random structure (SQS) approach [38], implemented in the Alloy Theoretic Automated Toolkit (ATAT) [39], was utilized to mimic the chemical disorder in binary diborides and HEBs with 4-9 TM elements (4-9HEBs). Nine TM elements were selected from group IVB, VB, and VIB, i.e., Ti, V, Cr, Zr, Nb, Mo, Hf, Ta, and W. All the structures were fully relaxed before the evaluation of the elastic tensor in DFT calculations by the energy-strain approach. The bulk modulus (*B*) and the shear modulus (*G*) of HEBs were estimated by the Voigt-Reuss-Hill method [40]. The *ab initio* molecular dynamics (AIMD) calculations were carried out with merely Γ point considered in the Brillouin zone to generate different configurations for self-consistent calculations. A timestep of 3 fs with up to 2000 steps was utilized in the NPT ensemble at 1000, 2000, and 3000 K, respectively.

The MTP was constructed using the Machine Learning Interatomic Potentials (MLIP-3) packages [41] with structural information on energies, forces, and stresses from DFT calculations. The training datasets were built according to our previously reported strategy [33], which merely required a collection of unary and binary diboride configurations with nine TM elements from AIMD. As there are nine compositions in unary diborides and 36 compositions in equimolar binary diborides, a total of 45 kinds of diborides were considered in the training dataset. For each composition at each temperature, 25 configurations were picked for training (a total of 3375 configurations) and five configurations for testing (a total of 675 configurations). Meanwhile, 50 equimolar and 50 non-equimolar HEBs with varied compositions were randomly generated, with two configurations collected from each to build equimolar and non-equimolar testing datasets (300 configurations for each dataset), respectively. To ensure the accuracy and efficiency of the MTP, the cutoff radius ($R_{cut}$) and the maximum level ($lev_{max}$) [42] were set to 6.5 Å and 22, respectively.

The MD simulations were carried out in the Large-scale Atomic/Molecular Massively Parallel Simulator (LAMMPS) package [43], interfaced with our trained MTP. All simulated HEB models included 3000 atoms, with cationic sites randomly occupied by TM elements with corresponding ratios. The timestep in all MD simulations was set to 1 fs, and the damping parameters for temperature and pressure were set to 0.1 ps and 1 ps, respectively. Equilibrium lattice constants were computed after 20000 steps of simulation in the NPT ensemble under 300 K, and the temperature-dependent elastic tensors were computed by the evaluation of the Born matrix under 300 K [44]. $T_m$ was simulated from 1500 K to 4500 K with a heating rate of 7.5 K/ps using the NPT ensemble after equilibrating at 1500 K using the NVT ensemble for 20 ps, and the coefficients of thermal expansion (CTEs) were obtained by evaluating equilibrium lattice constants from 300 K to 800 K with an interval of 100 K.

To identify $T_m$ of HEBs, six commonly used ML models, including linear regression (LR), support vector regression (SVR), decision tree regression (DTR), random forest regression (RFR), gradient boosting regression (GBR), and extreme gradient boosting (XGBoost), were adopted with the default hyperparameters [45]. The training dataset was collected with $T_m$ of 120 randomly selected HEBs and nine diborides predicted from MD simulations with our trained MTP (see **Table S1**) and 24 descriptors based on the fundamental physical parameters of constituent TM elements and diborides (see **Table 1**). The lattice size of diborides was the averaged lattice constants ($a$ and $c$) as they were comparable in values. $\overline{pam}$ and $\delta_{pam}$ stand the mean and difference values of fundamental parameters, respectively, and were computed as follows:

$$\overline{pam} = \sum_{i=1}^{n} x_i \, pam_i \tag{1}$$

$$\delta_{pam} = \sqrt{\sum_{i=1}^{n} x_i \left[\left(1 - \frac{pam_i}{\sum_{i=1}^{n} x_i \, pam_i}\right)^2\right]} \tag{2}$$

where $n$ is the number of constituent TM elements and diborides, and the $pam_i$ and $x_i$ correspond to the parameter and ratio of the $i$-th constituent TM element or diboride, respectively. The fundamental parameters of TM elements and diborides are summarized in **Table S2** and **S3**, respectively. To ensure comparability, descriptors were normalized to the same magnitudes by the following equations:

$$X_i^{\mathrm{norm}} = \frac{X_i - X_{\min}}{X_{\max} - X_{\min}} \tag{3}$$

where the $i$-th normalized value of a descriptor is denoted as $X_i^{\mathrm{norm}}$, and $X_i$ is the $i$-th value of the descriptor. $X_{\max}$ and $X_{\min}$ represent the maximum and the minimum values of the corresponding descriptor. The performance of ML models was estimated by the RMSE via the leave-one-out cross-validation approach [6].

**Table 1** Detailed 24 descriptors.

| Descriptors | Abbreviations | Descriptors | Abbreviations |
|---|---|---|---|
| Average valence electron concentration | $\overline{\mathrm{VEC}}$ | Valence electron concentration difference | $\delta_{\mathrm{VEC}}$ |
| Average atomic number | $\bar{N}$ | Atomic number difference | $\delta_N$ |
| Average effective nuclear charge | $\overline{Z^*}$ | Effective nuclear charge difference | $\delta_{Z^*}$ |
| Average Pauling electronegativity | $\overline{\chi^{\mathrm{P}}}$ | Pauling electronegativity difference | $\delta_{\chi^{\mathrm{P}}}$ |
| Average Mulliken electronegativity | $\overline{\chi^{\mathrm{M}}}$ | Mulliken electronegativity difference | $\delta_{\chi^{\mathrm{M}}}$ |
| Average Allen electronegativity | $\overline{\chi^{\mathrm{A}}}$ | Allen electronegativity difference | $\delta_{\chi^{\mathrm{A}}}$ |
| Average volume | $\bar{V}$ | Volume difference | $\delta_V$ |
| Average density | $\bar{\rho}$ | Density difference | $\delta_\rho$ |
| Average mass | $\bar{m}$ | Mass difference | $\delta_m$ |
| Average metallic radius | $\overline{r_{\mathrm{Me}}}$ | Metallic radius difference | $\delta_{r_{\mathrm{Me}}}$ |
| Average ionic radius | $\overline{r_{\mathrm{ion}}}$ | Ionic radius difference | $\delta_{r_{\mathrm{ion}}}$ |
| Average lattice size | $\bar{l}$ | Lattice size difference | $\delta_l$ |

GAs were employed to efficiently search the optimal combination of descriptors. The presence or absence of each descriptor was labeled as "1" or "0", respectively. In each GA run, an initial population of 100 descriptor combinations was randomly generated from 24 descriptors. The performance of each combination was estimated through the ML models, and then the "parents" were generated from the top 80% combinations. Subsequently, the new generation population was obtained via three main operators, namely selection, crossover, and mutation. The loop continued until it reached 50 iterations.

## 3. Results and discussion

The performance of the trained MTP was first evaluated. As depicted in **Fig. 2**(a,b), RMSEs of the energy and force predictions for both training and testing datasets consisting of unary and binary diborides are low, being approximately 2.6 meV/atom for energy and 158-161 meV/Å for force, respectively, suggesting the well-trained MTP. Then, the applicability of our constructed MTP to HEBs was examined. As shown in Fig. 2(c,d)**,** the testing accuracy of our constructed MTP for equimolar HEBs remains high, with almost no changes in RMSEs of energy and force (2.6 meV/atom and 155 meV/Å, respectively) compared to those of training datasets in Fig. 2(a,b). It should be noted that the performance of our trained MTP for non-equimolar HEBs is still remarkable, with low RMSEs comparable to those of the equimolar HEB testing dataset and the training dataset, achieving 3.7 meV/atom and 172 meV/Å for energy and force, respectively. These results demonstrate the generalization performance of our constructed MTP for both equimolar and non-equimolar HEBs with high accuracy and transferability. Moreover, the accuracy of our constructed MTP is higher than those of the previously reported MLPs for HECs and alloys f31,32,46], even with such a small volume of the training dataset, further showcasing the efficiency and accuracy of our MLP construction strategy. Based on these observations, it can be concluded that our trained MTP for HEBs is well-established, exhibiting high efficiency, accuracy, and transferability.

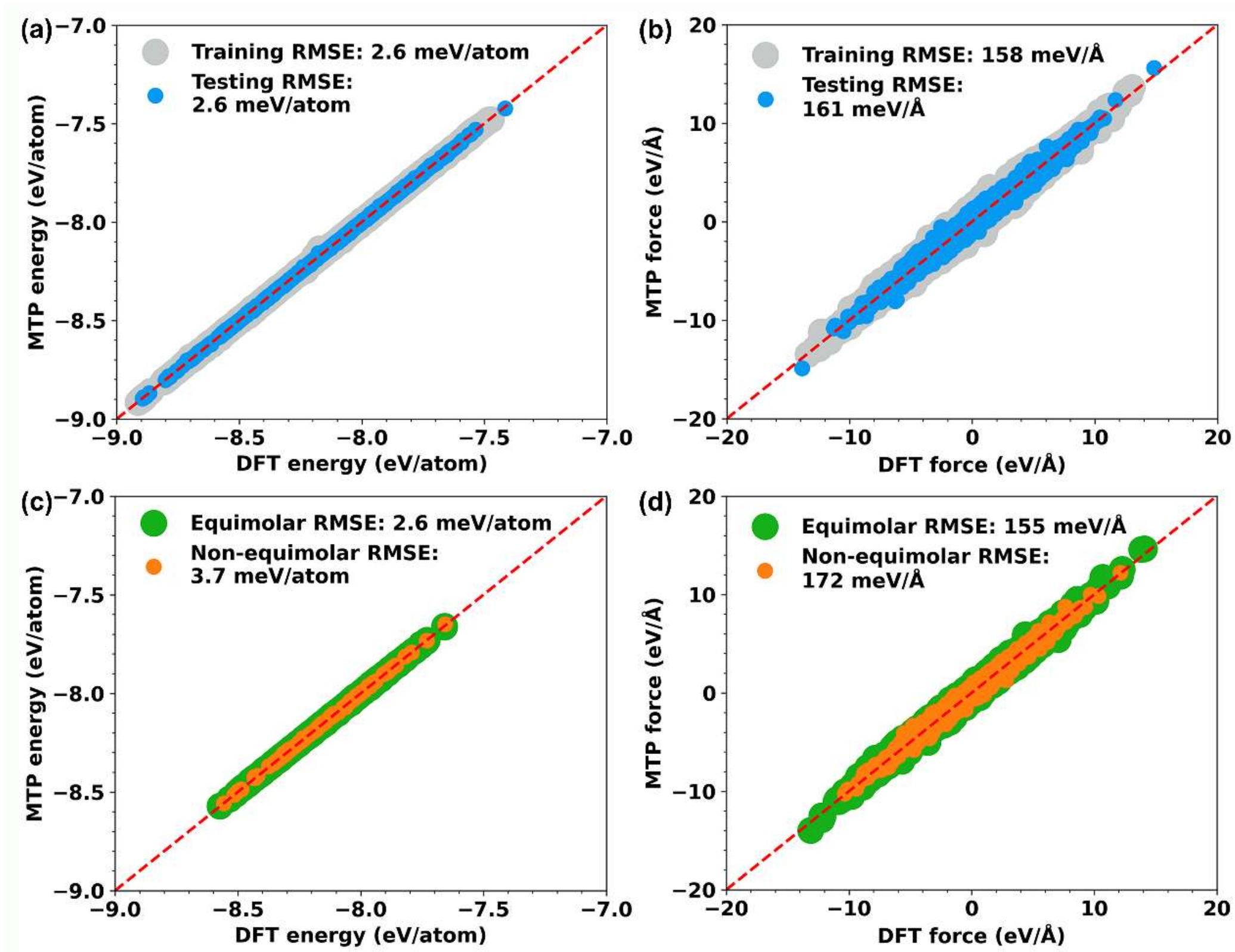

**Fig. 2** Performance of the constructed MTP compared with DFT calculations. (a) Energy and (b) force of the unary and binary diboride training and testing datasets. (c) Energy and (d) force of the equimolar and non-equimolar HEB testing datasets.

To further investigate the applicability of our constructed MTP, MLP-based MD simulations on structural and mechanical properties of HEBs were conducted. **Fig. 3** presents a comparative analysis between MD simulations with our trained MTP and DFT calculations as well as reported experiments [2,47-49], focusing on lattice parameters, elastic tensor ($c_{11}$, $c_{12}$, $c_{13}$, $c_{33}$, and $c_{44}$), $B$, and $G$ within a total of 50 randomly selected 4-9HEBs. It is found that $a$ and $c$ predicted from MD simulations with our well-trained MTP are in good agreement with those derived from DFT calculations (see Fig. 3(a,b)), indicating the accuracy and reliability of our constructed MTP. Comparable results between MD and DFT calculations can also be observed in elastic tensors from Fig. 3(c-g). The presence of small discrepancies between them originates from the geometric and method differences, where larger models and temperature effects were considered in our MD simulations. Similar results can also be obtained in Fig. 2(h,i), where *B* and *G* are slightly overestimated from DFT calculations compared with MLP-based MD simulations. Consequently, the reliability and accuracy of MD simulations with our well-constructed MTP for the properties of HEBs are validated.

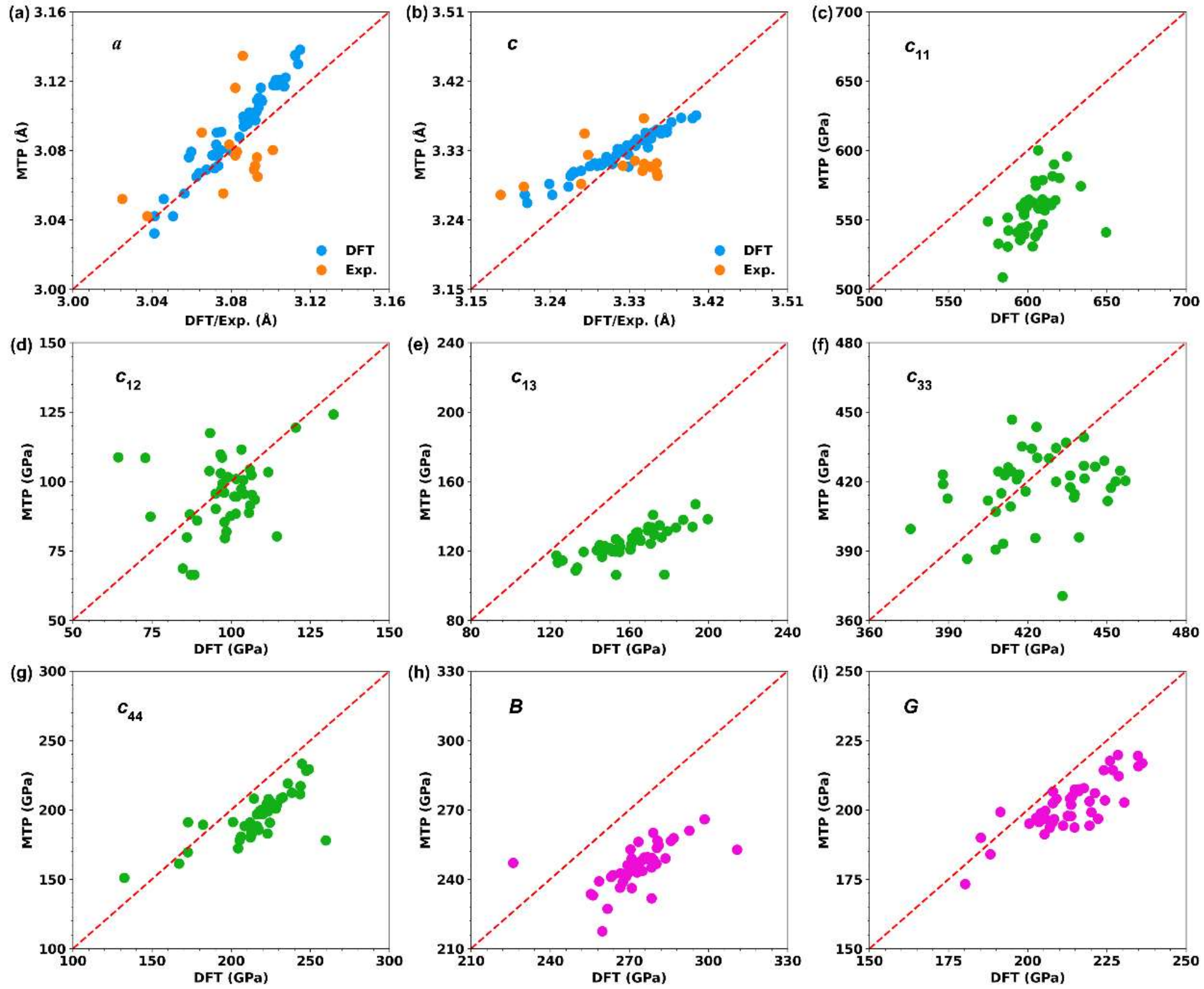

**Fig. 3** Comparison of structural and mechanical properties of 4-9HEBs between MD simulations with our fitted MTP and DFT calculations. (a, b) Lattice constants. Experimental data are from Ref. [2,47-49]. (c-g) Elastic tensors. (h) Bulk modulus. (i) Shear modules.

As the lack of available $T_m$ data for HEBs, the performance of MD simulations with our constructed MTP on predicting $T_m$ was compared with diborides first. It can be seen from **Fig. 4**(a) that all the predicted $T_m$ are well consistent with the experimental reports [50-52], implying the feasibility and accuracy of MD predictions with our constructed MTP in predicting $T_m$ of HEBs. Besides nine diborides, $T_m$ of 120 randomly selected equimolar and non-equimolar HEBs (listed in Table S1) were accurately predicted to ensure adequate data in the training dataset, and 24 descriptors were collected in the meantime according to their compositions. Different ML models with GAs were trained, and the results are summarized in **Table 2**. Among all six ML models, the GBR model is found to exhibit the best validation accuracy with the lowest RMSE of 93.8 K. The proposed combination of descriptors contains five descriptors, namely $\overline{\mathrm{VEC}}$, $\delta_l$, $\delta_{r_{\mathrm{Me}}}$, $\bar{\rho}$, and $\delta_{\chi^{\mathrm{A}}}$. In addition, Fig. 4(b) presents the best validation accuracy of GBR models with different GA runs. It can be observed that the stable optimal validation accuracy occurs after 11 independent GA runs with RMSE changes less than 1 K. Fig. 4(c) plots the validation accuracy of GBR models as a function of iterations. The convergence of both the average and the maximum validation accuracies with the increase of iteration numbers in GBR models can be found, which further verifies the reliability of the established GBR model.

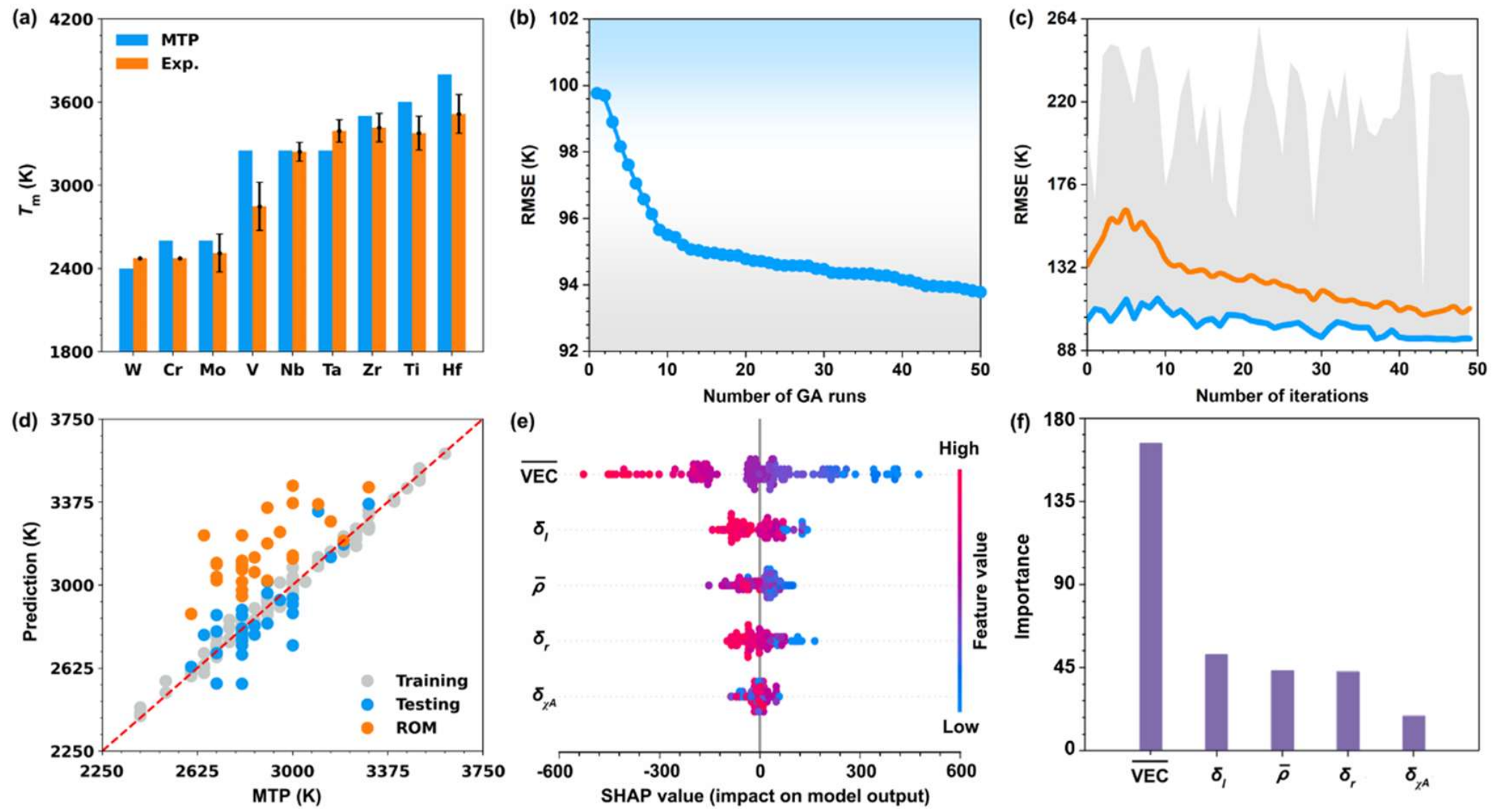

**Fig. 4** Performance of $T_m$ prediction from MD simulations with our trained MTP, ROM, and ML models. (a) $T_m$ of diborides from MD simulations with our trained MTP and reported experiments [50-52]. (b) Best validation accuracy of GBR models as a function of the GA run number. (c) Validation accuracy of GBR models as a function of iteration number. Orange and blue lines represent each iteration's best performance and mean accuracy, respectively. (d) Comparison of $T_m$ between MD simulations with our trained MTP and the established GBR model as well as ROM. (e) SHAP values for the training dataset. (f) Importance of descriptors.

**Table 2** RMSEs of different ML models after GA runs.

| ML Models | GBR | LR | SVR | DTR | RFR | XGBoost |
|---|---|---|---|---|---|---|
| RMSE (K) | 93.8 | 101.5 | 100.5 | 108.5 | 99.6 | 106.6 |

To further examine the accuracy of our established GBR model, the performance on a testing dataset with 30 randomly selected equimolar and non-equimolar HEBs (summarized in **Table S4**) is presented in Fig. 4(d). A high testing accuracy with a low RMSE of 109.5 K is obtained, which is also significantly smaller than the one predicted from the rule of mixture (ROM; RMSE = 305.7 K). These observations show the reliability of GA runs in GBR models for $T_m$ predictions in HEBs and $\overline{\mathrm{VEC}}$, $\delta_l$, $\delta_{r_{\mathrm{Me}}}$, $\bar{\rho}$, and $\delta_{\chi^{\mathrm{A}}}$ are confirmed as the optimal descriptor combination. To interpret the ML models, Shapley additive explanation (SHAP) was used. As the SHAP values and the importance (mean absolute SHAP values) illustrated in Fig. 4(e,f), $\overline{\mathrm{VEC}}$ is identified to be the most important descriptor with a negative correlation in $T_m$, followed by $\delta_l$, $\bar{\rho}$, $\delta_{r_{\mathrm{Me}}}$, and $\delta_{\chi^{\mathrm{A}}}$. Therefore, a lower $\overline{\mathrm{VEC}}$ is preferable to realize high $T_m$ in HEBs. Additionally, smaller $\delta_l$, $\bar{\rho}$, and $\delta_{r_{\mathrm{Me}}}$ are also beneficial for achieving high $T_m$ according to our established GBR model.

With the established optimal combination of descriptors and the GBR model, high-throughput $T_m$ predictions on equimolar and non-equimolar HEBs can be efficiently performed. **Fig. 5** demonstrates the distribution of predicted $T_m$ in both equimolar (382 compositions) and non-equimolar 4-9 HEBs (32181 compositions). It can be seen that although there are no candidates possessing $T_m$ outperforming the maximum of diborides ($HfB_2$, 3800 K), a number of potential HEBs with $T_m$ larger than 3273 K have been found. Moreover, HEBs with non-equimolar and fewer TM elements tend to possess broader $T_m$ ranges, leading to more ultra-high $T_m$ HEBs observed in these domains. Consequently, a total of 2284 HEBs with $T_m$ larger than 3273 K were predicted, where 14 are in 4-6HEB equimolar systems from Fig. 5(a) and 2270 are in 4-7HEB non-equimolar systems from Fig. 5(b). Detailed potential ultra-high temperature ceramics (UHTCs) are summarized in **Table S5** and **S6**. It can also be found that TiZrHf-based HEBs tend to be promising UHTCs, achieving the maximum $T_m$ of approximately 3688 K in the non-equimolar $(Ti_{0.1}Zr_{0.1}Hf_{0.6}Ta_{0.2})B_2$ system.

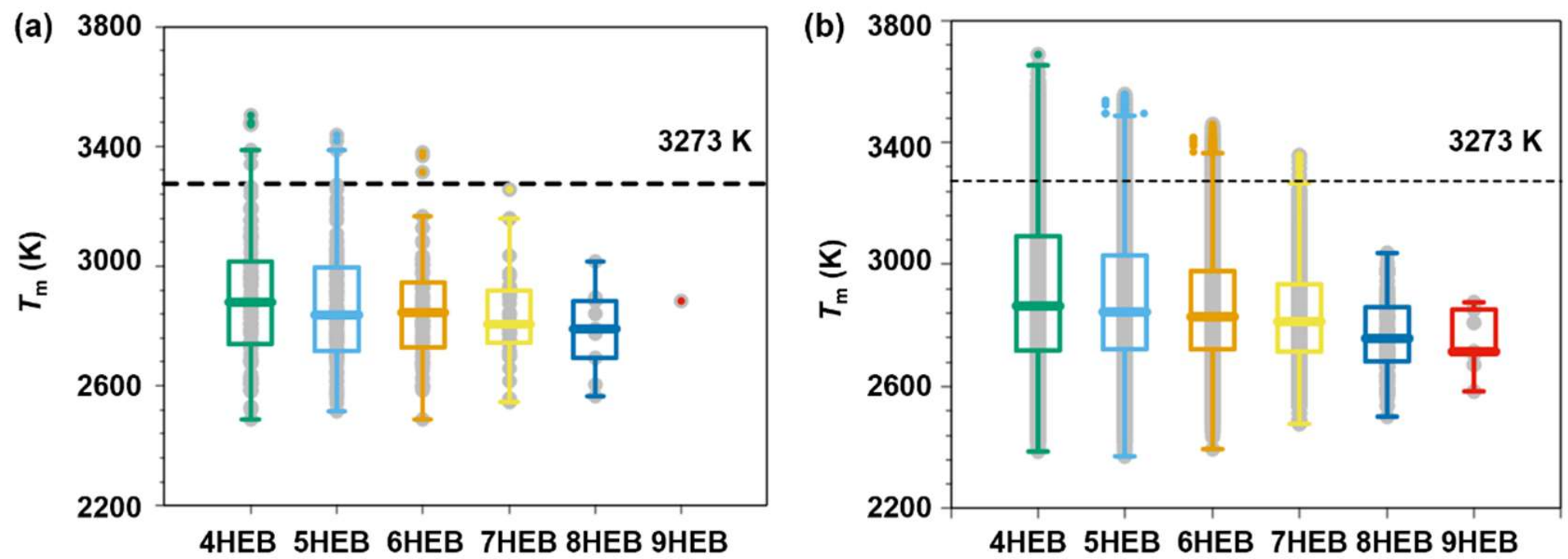

**Fig. 5** Distribution of predicted $T_m$ in different HEBs from our established GBR model. (a) Equimolar 4-9HEBs. (b) Non-equimolar 4-9HEBs.

In addition to $T_m$, the MLP-driven prediction can also be utilized to investigate many other properties of HECs. For example, **Fig. 6**(a) shows the CTEs of diborides from MD simulations with our constructed MTP and the corresponding reported experimental values [53-60]. Similar to $T_m$, the results of MD are in alignment with experimental reports with relatively low discrepancies, indicating the accuracy of our MLP-based MD predictions in HEBs' CTEs. Fig. 6(b) shows the comparison of CTEs in 50 randomly selected 4-9HEB between MD simulations with our constructed MTP and ROM. Clearly, predictions of CTEs based on our constructed MTP are more accurate than ones from ROM, which tends to be underestimated. Although the MD predictions of CTEs with our well-fitted MTP are feasible, the computational intensity still hinders its high-throughput exploration in CTEs of HEBs. Therefore, it is urgent to build the CTEs training datasets with the aid of the MTP to enable further ML predictions in HEBs with desired CTEs.

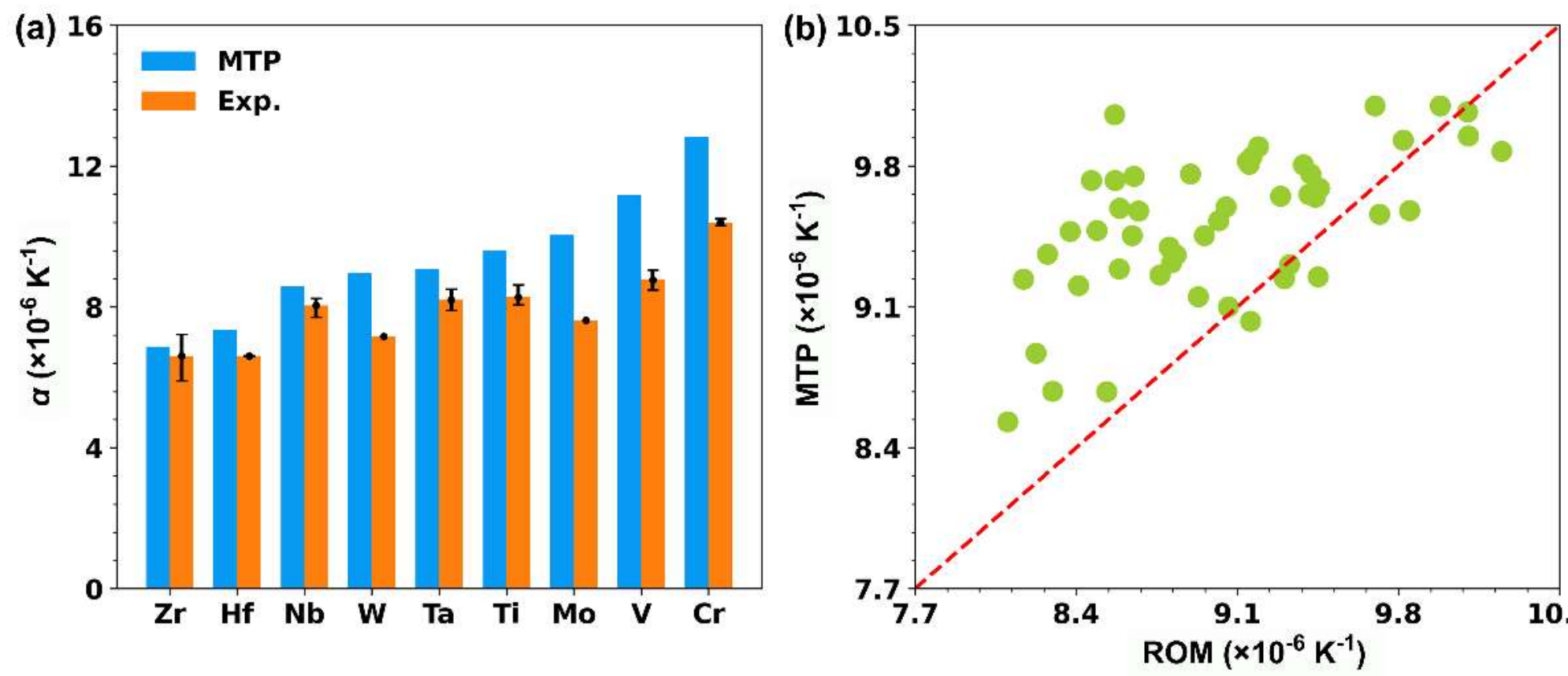

**Fig. 6** CTEs of diborides and 4-9HEBs from MD simulations with our trained MTP, experimental measurements, and ROM calculations. (a) CTEs of different diborides from MD simulations with our trained MTP and experimental reports [53-60]. Error bars are the maximum and minimum of the reported values. (b) CTEs of different 4-9HEBs between MD simulations with our trained MTP and ROM calculations.

## 4. Conclusion

In summary, taking HEBs as an example, a data-driven approach has been developed to high-throughput search for HEBs with ultra-high $T_m$ via MD predictions based on our transferable MTP. Specifically, an efficient, accurate, and transferable MTP with broad compositional applicability (up to nine TM elements) for HEBs was well-constructed from unary and binary diborides. Low RMSEs of energy and force in both equimolar and non-equimolar HEB testing datasets demonstrated the remarkable accuracy and transferability of our constructed MTP. The comparison of the structural and mechanical properties between MD simulations with the trained MTP and DFT calculations further revealed the reliability of our trained MTP. The accurately simulated $T_m$ of 120 HEBs and nine diborides from MD simulations with our constructed MTP were collected with 24 descriptors to achieve reliable ML models for $T_m$ prediction in HEBs. A combination of five descriptors ($\overline{\mathrm{VEC}}$, $\delta_l$, $\delta_{r_{\mathrm{Me}}}$, $\bar{\rho}$, and $\delta_{\chi^{\mathrm{A}}}$) were identified as the optimal descriptor combination by ML approaches combined with GAs. With the established descriptors, $T_m$ of 32563 HEBs were predicted, where the largest $T_m$ of up to 3688 K in the non-equimolar $(Ti_{0.1}Zr_{0.1}Hf_{0.6}Ta_{0.2})B_2$. Our work provides an efficient MLP-driven approach to search for HECs with ultra-high $T_m$, and is also applicable to exploring HECs with other desired properties.

## Acknowledgements

We acknowledge the financial support from the National Key Research and Development Program of China (No. 2021YFA0715801), the National Natural Science Foundation of China (No. 52122204 and 52402075), and Guangzhou Basic and Applied Basic Research Foundation (SL2023A04J00690).

## Author Contributions

Y. Chu conceived and designed this work. H. Meng, Y. Liu, and H. Yu performed the theoretical calculations. Y. Chu, H. Yu, H. Meng, and L. Zhuang analyzed the data and wrote the manuscript. All authors commented on the manuscript.

## Competing Interests

The authors declare no competing financial interest.

## Data Availability

The data that support the findings of this study are available from the corresponding author upon reasonable request.